\documentclass[aps,prl,preprint,groupedaddress]{revtex4-2}
\usepackage{graphicx}
\usepackage{dcolumn}%
\usepackage{bm}
\usepackage[version=3]{mhchem}
\usepackage[left=1.5cm, right=1.5cm, top=1.785cm, bottom=2.0cm]{geometry}
\usepackage{balance}
\usepackage{amssymb}
\usepackage{mathptmx}
\usepackage{graphicx} 
\usepackage{lastpage}
\usepackage{color}
\usepackage{soul}
\usepackage[format=plain,justification=justified,singlelinecheck=false,font={stretch=1.125,small,sf},labelfont=bf,labelsep=space]{caption}
\usepackage{float}
\usepackage{fancyhdr}
\usepackage{fnpos}
\usepackage[english]{babel}
\addto{\captionsenglish}{
 
}
\usepackage{array}
\usepackage{droidsans}
\usepackage{charter}
\usepackage[T1]{fontenc}
\usepackage[usenames,dvipsnames]{xcolor}
\usepackage{setspace}
\usepackage[compact]{titlesec}
\usepackage{hyperref}

\usepackage{epstopdf}

\usepackage[normalem]{ulem}
\usepackage{framed}

\definecolor{cream}{RGB}{222,217,201}

\begin{document}

\title{Geometric fields and new enantio-sensitive observables in photoionization of chiral molecules }

\author{Andres F. Ordonez$^{1,2}$, David Ayuso$^{1,3}$, Piero Decleva$^{4,5}$ and Olga Smirnova$^{1,6}$}

\affiliation{
$^1$Max-Born-Institut, Max-Born-Str. 2A, 12489 Berlin, Germany\\
$^2$ICFO-Institut de Ciencies Fotoniques, The Barcelona Institute of Science and Technology, 08860 Castelldefels (Barcelona), Spain\\
$^3$Department of Physics, Imperial College London, SW7 2BW London, United Kingdom\\
$^4$Dipartimento di Scienze Chimiche e Farmaceutiche, Universit\`a degli Studi di Trieste, Trieste, Italy\\
$^5$CNR-IOM DEMOCRITOS, Trieste, Italy\\
$^6$Technische Universit\"at Berlin, Straße des 17. Juni 135, 10623 Berlin, Germany}

\date{\today}

\begin{abstract}
Chiral molecules are instrumental for molecular recognition in living organisms.
Distinguishing between two opposite enantiomers, the mirror twins of the same chiral molecule, is both vital and challenging.
Photoelectron circular dichroism (PECD), an extremely sensitive probe of molecular chirality via photoionization, outperforms standard optical methods by many orders of magnitude. Here we show that the physical origin of PECD in chiral molecules is linked to the concept of geometric magnetism, which enables a broad class of phenomena in solids including the anomalous electron velocity, the Hall effect, and related topological phenomena. We uncover the geometric field in molecular photoionization, which leads to a new class of enantio-sensitive observables emerging due to ultrafast excitation of chiral electronic or vibronic currents prior to ionization. Next, we introduce the first member of this new class: enantio-sensitive orientation of chiral molecules via photoionization.
This effect opens new routes to both enantio-separation and imaging of chiral dynamics on ultrafast time scales.
Our work suggests that geometric fields in photoionization provide the bridge between the two geometrical properties, chirality and topology.
\end{abstract}

\maketitle

Chiral molecules are characterised by their handedness, an extra degree of freedom of a purely geometrical origin. Geometrical properties in real space are determined by the nuclear configuration in molecules or by the lattice configuration in solids. In solids, they map onto geometrical or topological properties of Hilbert space vectors, leading to robust observables associated with the electronic response to electromagnetic fields and new, topological, phases of matter. Geometric magnetism, introduced by M. Berry \cite{berry1984quantal}, is a key concept underlying these phenomena. One of the manifestations of geometric magnetism is the Berry curvature in solids, which enables a class of new observables in condensed matter systems related to the so-called anomalous electron velocity imparted by the Berry curvature \cite{RevModPhys.66.899, RevModPhys_Xiao_2010}.

We show that similar geometric field and ``anomalous'' observables induced by such field also arise in one\cite{Ritchie1976PRA,Powis2000JCP,Bowering2001PRL,Nahon2015JESRP,Janssen2014PCCP}-- or multiphoton\cite{Lux2012Angewandte,Lehmann2013JCP,beaulieu2016universality} ionization of randomly oriented chiral molecules by circularly polarized fields, with photoelectron circular dichroism (PECD)\cite{Ritchie1976PRA,Powis2000JCP,Bowering2001PRL} being one example of such ``anomalous'' phenomena. The PECD signal is encoded in the direction of the net photoelectron current, which is opposite in opposite enantiomers and is perpendicular to the light polarization plane. Unlike standard absorption circular dichroism (CD), PECD arises already in the electric-dipole approximation and is associated with very strong enantio-sensitive signals.

We show that not only the emergence and enantio-sensitivity of PECD is linked to the concept of geometric field, but that this concept also allows one to predict new efficient enantio-sensitive observables. One of them is enantio-sensitive orientation of randomly oriented chiral molecules via two- or multi-photon ionization, which we call ``molecular orientation circular dichroism'' (MOCD). Crucially, these new observables rely on ultrafast excitation of chiral electronic or vibronic currents, linking geometric fields to yet another important concept in ultrafast science: the concept of charge-directed reactivity.

Charge directed chemical reactivity \cite{weinkauf1997nonstationary,Cederbaum1999,Breidbach2003,Remacle2006,Kuleff2014,Calegari2014,Nisoli2017,D0FD00121J} implies that ultrafast electron dynamics can affect the outcome of chemical reactions, opening an important direction in attochemistry \cite{corkum1997subfemtosecond,Cederbaum1999,Breidbach2003,Remacle2006,Kuleff2014,Calegari2014,Nisoli2017,D0FD00121J}. We show that ultrafast electron currents can lead to opposite orientation of left and right enantiomers of the same molecule upon photoionization of randomly oriented molecules by circularly polarized light, thus presenting an example of enantio-sensitive charge-directed reactivity, with geometric fields and concepts providing a platform for its description. 

\section{Geometric field: molecules vs solids}

Geometric concepts in photoionizaton of chiral molecules with circularly polarized light originate from to the so-called propensity field, which we have introduced recently \cite{ordonez_propensity_2019}. It emerges in photoionization of randomly oriented molecules and involves the vector product of two conjugated photoionization dipoles ($\vec{d}_{\vec{k}g}$ in the length or $\vec{p}_{\vec{k}g}$ in the velocity gauges): 
\begin{equation}
 \vec{B}(\vec{k})=i[\vec{d}_{\vec{k}g}\times \vec{d}^*_{\vec{k}g}]=i\frac{[\vec{p}_{\vec{k}g}\times \vec{p}^*_{\vec{k}g}]}{(E_k-E_g)^2}.
\label{eq:fieldB}
\end{equation}
Here $E_k=k^2/2$ is the photoelectron energy and $E_g$ is the energy of the initial (e.g. ground) state. The field $\vec{B}(\vec{k})$ is a molecular-frame property.
 
The second equality in Eq. (\ref{eq:fieldB}) is also true for the Berry curvature in a two-band solid \cite{Yao_valley_2008}:
 \begin{equation}
 \vec{\Omega}(\vec{k})=i\frac{[\vec{p}^{vc}_{\vec{k}}\times \vec{p}^{vc*}_{\vec{k}}]}{(E_c-E_v)^2},
\label{eq:fieldBer}
\end{equation}
where the matrix elements $\vec{p}^{vc}$ and energies $E_{v,c}$ describe transitions between the valence ($v$) and conduction ($c$) bands. Just like the Berry curvature (Eq. (\ref{eq:fieldBer})) characterises $\vec{k}$-dependent circular dichroism in interband transitions in solids\cite{Yao_valley_2008}, the propensity field (Eq. (\ref{eq:fieldB})) characterises $\vec{k}$-dependent circular dichroism in photoionization\cite{ordonez_propensity_2019}.

The analogy between $\vec{B}(\vec{k})$
 [Eq. (\ref{eq:fieldB})] and $\vec{\Omega}(\vec{k})$ 
[Eq. (\ref{eq:fieldBer})] runs deep. 
The enantio-sensitive photoionization current injected per unit time, describing PECD in molecules, is orthogonal to the polarization plane of the driving laser light, and can be written via the enantio-sensitive and $k$-dependent conductivity $\sigma_{zz}(k)$ as:
\begin{eqnarray}
\label{eq:PECD_current}
\frac{d{j}_z}{dt}=\sigma_{zz} [\vec{E}_{\omega}\times\vec{E}^*_{\omega}]_z,\label{eq:currentB}
\end{eqnarray}
where\cite{ordonez2019propensity2} $\sigma_{zz}(k)$ is proportional to the flux of the field $\vec{{B}}(\vec{k})$ [Eq. (\ref{eq:fieldB})] through the sphere in $\vec{k}$-space with radius $k=\sqrt{2E_k}$ and the surface element $d\vec{S}=k^2d\Theta_{k} (\vec{k}/k)$, where $d\Theta_{{k}}\equiv d\phi_kd\theta_k\sin\theta_k$:
\begin{eqnarray}
\sigma_{zz}(k)=\frac{1}{6k}\oint \vec{{B}}(\vec{k})\cdot d\vec{S}.\label{eq:conductB}
\end{eqnarray}

The solid state analogue to PECD is the circular photogalvanic effect in isotropic enantiomorphic (chiral) solids: circularly polarized light induces net electron current perpendicular to its polarization plane. This current has recently been related \cite{Juan2017quantized} to the Berry curvature $\vec{\Omega}(\vec{k})$ as\footnote{Note that Juan et al. \cite{Juan2017quantized} use $\beta=-i\sigma$ to formulate Eqs.\ref{eq:currentBer},\ref{eq:conductBer}}
 \begin{equation}
\frac{d{j}_z}{dt}=\sigma_{zz} [\vec{E}_{\omega}\times\vec{E}^*_{\omega}]_z,\label{eq:currentBer}
\end{equation}
\begin{equation}
\sigma_{zz}(k)=\frac{e^3}{6h^2}\oint \vec{\Omega}(\vec{k})\cdot d\vec{S}.
\label{eq:conductBer}
\end{equation}
Eq. (\ref{eq:conductBer}) is exact in the two-band approximation. The similarity between Eqs. (\ref{eq:currentBer},\ref{eq:conductBer}) and (\ref{eq:currentB},\ref{eq:conductB}) is clear. 

 Using Stokes theorem Eq. (\ref{eq:conductB}) can be rewritten via the volume integral of the divergence $\vec{\nabla}_{k}\cdot \vec{B}(\vec{k})=h(\vec{k})$, where the pseudoscalar quantity $h(\vec{k})$ unique to chiral systems is the enantio-sensitive helicity of the photoionizaion dipole: $h(\vec{k})\equiv\Im\{i\vec{d}_{\vec{k}g}^{*}\cdot\vec{\nabla_k} \times\vec{d}_{\vec{k}g}\}$. Thus, $\vec{B}(\vec{k})$ reflects the geometry (helicity) of the molecular photoionization dipoles.

\begin{figure*}[h]
\includegraphics[width=\textwidth]{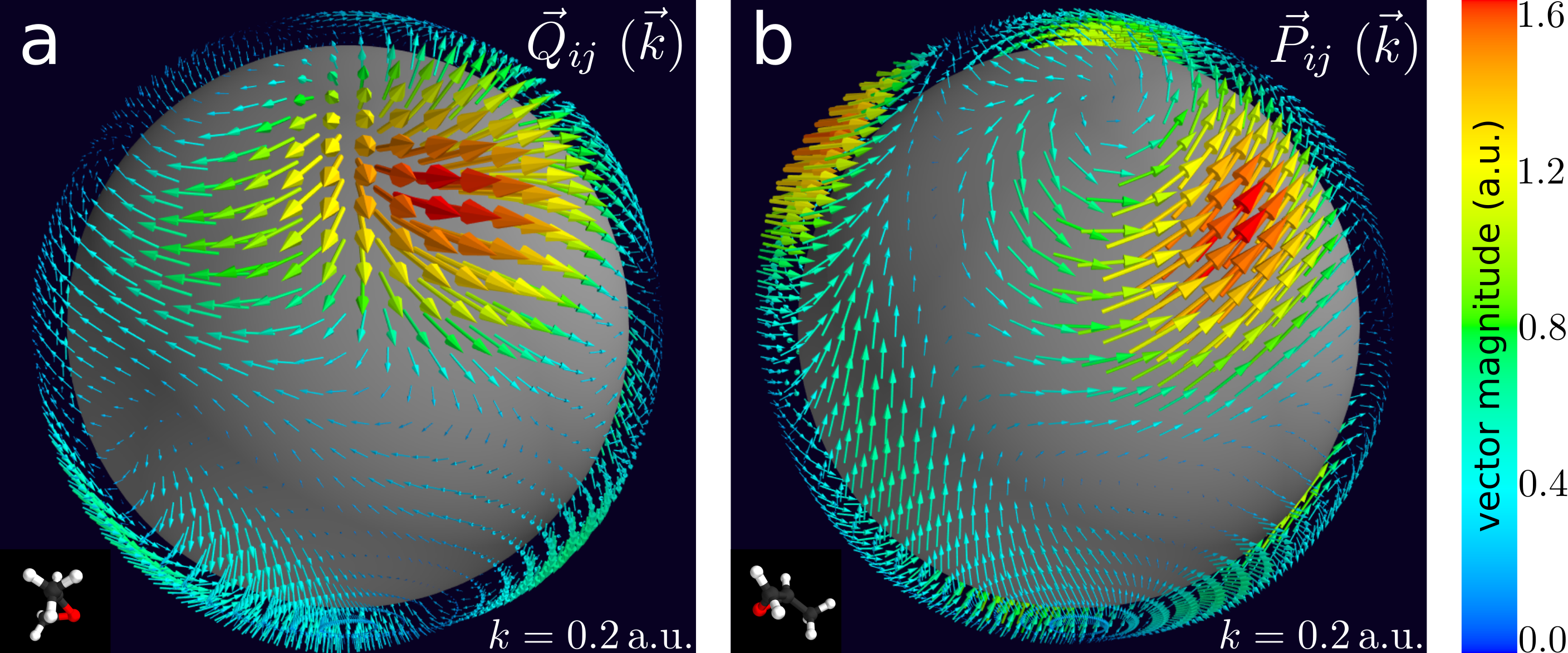}
\caption{\label{fig:new} (a) Displacement $\vec{Q}_{ij}(\vec{k})$ [Eq. (\ref{eq:Q_ij_definition})] and (b) current $\vec{P}_{ij}(\vec{k})$ [Eq. (\ref{eq:P_ij_definition})] quadratures of the geometric field $\vec{B}_{i,j}(\vec{k})$ [Eq. (\ref{eq:B_ij})] for $i$=LUMO and $j$=LUMO+1 of the chiral molecule propylene oxide and photoelectron momentum $k=0.2$ a.u. The molecular orientation
is shown in the left bottom corner of each panel. Note that the quadratures are shown from different viewpoints.}
\end{figure*}

 In photoionization from a superposition of two states $|j\rangle+e^{-i\phi_{ij}}|i\rangle$ ($\phi_{ij}=\omega_{ij}t$, with $\omega_{ij}\equiv\omega_i-\omega_j$ being the transition frequency between the states) the 
propensity field acquires an additional term, which encodes the coherence between the excited states:
\begin{equation}
\label{eq:B_ij}
\vec{B}_{ij}(\vec{k},\phi_{ij})=-\frac{1}{2}i\left[\vec{d}^{*}_{\vec{k}i}\times \vec{d}_{\vec{k}j}\right]e^{i\phi_{ij}}+\mathrm{c.c.}\equiv\vec{Q}_{ij}(\vec{k})\cos\phi_{ij}+\vec{P}_{ij}(\vec{k})\sin\phi_{ij},
\end{equation}
where we have introduced the displacement $\vec{Q}_{ij}(\vec{k})$ and current $\vec{P}_{ij}(\vec{k})$ quadratures: 
\begin{equation}
\vec{Q}_{ij}(\vec{k})\equiv -\Re\left\{i\left[\vec{d}^{*}_{\vec{k}i}\times \vec{d}_{\vec{k}j}\right]\right\},
\label{eq:Q_ij_definition}
\end{equation}
\begin{equation}
\vec{P}_{ij}(\vec{k})\equiv \Im\left\{i\left[\vec{d}^{*}_{\vec{k}i}\times \vec{d}_{\vec{k}j}\right]\right\}.
\label{eq:P_ij_definition}
\end{equation}
Examples of the displacement $\vec{Q}_{ij}(\vec{k})$ and current $\vec{P}_{ij}(\vec{k})$ quadratures for a coherent superposition of $i$=LUMO and $j$=LUMO+1 in propylene oxide (a chiral molecule) are shown in Fig. \ref{fig:new}.

Eq. (\ref{eq:B_ij}) describes the component of the field oscillating at the frequency $\omega_{ij}$. For $i=j=g$, $\phi_{ij}=0$ and
Eq. (\ref{eq:B_ij}) reduces to Eq. (\ref{eq:fieldB}). For any number of states Eqs. (\ref{eq:fieldB},\ref{eq:B_ij}) can be generalized as
\begin{equation}
\label{eq:B_ij_sum}
\vec{B}(\vec{k},t)=\frac{1}{2}\sum_{i,j}\left\{i\left[\vec{d}_{\vec{k}j}\times \vec{d}^{*}_{\vec{k}i}\right]\right\}e^{i\phi_{ij}}.
\end{equation}

We call $\vec{B}(\vec{k},t)$ in Eq. (\ref{eq:B_ij_sum}) \emph{geometric field in molecular photoionization}. Applying inversion ($\vec{r}\rightarrow -\vec{r}$, $\vec{k}\rightarrow -\vec{k}$) to reverse molecular handedness, we find that the displacement and current quadratures in left- (S) and right-handed (R) molecules are connected via $\vec{Q}^{(S)}_{ij}(\vec{k})=\vec{Q}^{(R)}_{ij}(-\vec{k})$ and $\vec{P}^{(S)}_{ij}(\vec{k})=\vec{P}^{(R)}_{ij}(-\vec{k})$. 

The geometric field Eq. (\ref{eq:B_ij_sum}) gives rise to three classes of enantio-sensitive observables in photoionization of randomly oriented molecules.

\section{Three classes of anomalous enantio-sensitive observables}
Enantio-sensitive photoionization observables are defined in the laboratory frame and originate from global invariants of the geometric field -- quantities surviving averaging over the directions of $\vec{k}$ in the molecular frame.

The \textbf{first} global invariant corresponds to non-zero \emph{net} geometric field (integrated over all angles $\phi_k,\theta_k$ characterizing the orientation of $\vec{k}$ in the molecular frame, 
$d\Theta_{{k}}\equiv d\phi_kd\theta_k\sin\theta_k$):
\begin{eqnarray}\label{eq:I1}
\vec{\mathsf{B}}_{ij}(k)\equiv\int \vec{B}_{ij}(\vec{k})d\Theta_k\neq0.
\end{eqnarray}

The \textbf{second} global invariant
requires non-zero \emph{net} radial component of the geometric field:
\begin{eqnarray}\label{eq:I2}
\left[\mathsf{B}_{\parallel}(k)\right]_{ij}\equiv\int \vec{B}_{ij}(\vec{k})\cdot\hat{k}d\Theta_k\neq0,\hspace{1.5 cm} \hat{k}=\frac{\vec{k}}{k}.
\end{eqnarray}

The \textbf{third} global invariant includes an infinite array of multipoles ($l\geq 1$) of \emph{net radial} and of the two \emph{net tangential} components of the geometric field:
\begin{equation}
\label{eq:I3multi_r}
\left[\mathsf{B}_{\parallel}^{l,m}(k)\right]_{ij}\equiv\int \vec{B}_{ij}(\vec{k})\cdot\hat{k}Y_{lm}\left(\theta_k,\phi_k\right)d\Theta_k\neq0, \hspace{1.5 cm} \hat{k}=\frac{\vec{k}}{k}
\end{equation}
\begin{equation}
\label{eq:I3multi_t1}
\left[\mathsf{B}_{\perp, 1}^{l,m}(k)\right]_{ij}\equiv \int \vec{B}_{ij}(\vec{k})\cdot\vec{\nabla}_k Y_{lm}\left(\theta_k,\phi_k\right)d\Theta_k\neq0,
\end{equation}
\begin{equation}
\label{eq:I3multi_t2}
\left[\mathsf{B}_{\perp, 2}^{l,m}(k)\right]_{ij}\equiv\int \vec{B}_{ij}(\vec{k})\cdot\left[\hat{k}\times\vec{\nabla}_k \right]Y_{lm}\left(\theta_k,\phi_k\right)d\Theta_k\neq0
\end{equation}
These quantities are the spherical multipole moments of the vector field $\vec{B}(\vec{k})$ (see e.g. \cite{barrera_vector_1985}).

These three global invariants lead to three classes of enantio-sensitive observables in photoionization of randomly oriented molecules. We call these observables anomalous, in analogy to anomalous velocity in solids stemming from the Berry curvature.
Their emergence or cancellation is determined by the time-reversal symmetry of the molecular bound states and the $\vec{k}$-inversion symmetries of the quadratures $\vec{Q}_{ij}(\vec{k}), \vec{P}_{ij}(\vec{k})$.
It is convenient to introduce symmetric and anti-symmetric superpositions of the quadratures corresponding to left ($\vec{Q}_{ij}^{(S)}, \vec{P}_{ij}^{(S)}$) and right ($\vec{Q}_{ij}^{(R)}, \vec{P}_{ij}^{(R)}$) molecules:
\begin{equation}
\vec{Q}^{\pm}_{ij}(\vec{k})=\frac{1}{2}\left[\vec{Q}^{(S)}_{ij}(\vec{k})\pm\vec{Q}^{(R)}_{ij}(\vec{k})\right],
\label{eq:Q_evenodd_definition}
\end{equation}
\begin{equation}
\vec{P}^{\pm}_{ij}(\vec{k})=\frac{1}{2}\left[\vec{P}^{(S)}_{ij}(\vec{k}){\pm}\vec{P}^{(R)}_{ij}(\vec{k})\right].
\label{eq:P_evenodd_definition}
\end{equation}

The symmetric superpositions $\vec{Q}^{+}_{ij}(\vec{k})$, $\vec{P}^{+}_{ij}(\vec{k})$ are $\vec{k}$-even and the anti-symmetric superpositions $\vec{Q}^{-}_{ij}(\vec{k})$, $\vec{P}^{-}_{ij}(\vec{k})$ are $\vec{k}$-odd.

The \textbf{Class I} of enantio-sensitive observables relies on the existence of the net geometric field Eq. (\ref{eq:I1}). Evidently, the $k$-odd quadratures $\vec{Q}_{ij}^{-}(\vec{k})$ and $\vec{P}_{ij}^{-}(\vec{k})$ do not contribute to this integral. We show (see Methods) that Eq. (\ref{eq:I1}) cannot be satisfied when ionization takes place from a real (time-even) state. Thus, $\vec{Q}_{ij}^{+}(\vec{k})$ also does not contribute to the net geometric field, which only arises due to the symmetric quadrature $\vec{\mathrm{P}}_{ij}^{+}(k)$:
\begin{eqnarray}
\label{eq:netB1}
\vec{\mathsf{B}}_{ij}(k)= \int \vec{B}_{ij}(\vec{k})d\Theta_k=\int \vec{P}_{ij}^{+}(\vec{k})d\Theta_k\sin\phi_{ij}\equiv\vec{\mathsf{P}}_{ij}^{+}(k)\sin\phi_{ij}
\end{eqnarray}
Thus, the new enantio-sensitive observables of Class I can only appear if photoionization by a circularly polarized field occurs from current-carrying states ($\sin\phi_{ij}\neq 0$). For example, such states can be generated by a pump pulse with ionization following ultrafast excitation of a coherent superposition of eigenstates. The fact that the net field $\vec{\mathrm{B}}(k)$ emerges only in systems undergoing dynamics makes it an important player in attosecond photochemistry. In contrast to ring currents excited in atoms or non-chiral molecules by circularly polarized fields \cite{Eckart2018,Barth2007PRA,Bandrauk}, chiral molecules present an example of a system where excited currents do not vanish in the molecular frame
upon averaging over random molecular orientations, because they are protected by the rotationally invariant geometric property of \emph{molecular handedness}. The net geometric field in the molecular frame (\ref{eq:netB1}) leads to enantio-sensitive molecular orientation by ionization (see below) and is an example of charge directed reactivity emerging solely due to molecular handedness.

The \textbf{Class II} of enantio-sensitive observables relies on the existence of the \emph{net radial} component of the geometric field Eq. (\ref{eq:I2}). Clearly, the symmetric quadratures $\vec{Q}^{+}(\vec{k})$ and $\vec{P}^{+}(\vec{k})$ do not contribute to $\mathsf{B}_{||}(k)$, leaving us with the following global invariants allowed by symmetries:
\begin{equation}
\label{eq:I2_jB}
\left[\mathsf{B}_{\parallel}(k)\right]_{ij}=\left[\mathsf{Q}_{\parallel}^{-}(k)\right]_{ij}\cos\phi_{ij}+\left[\mathsf{P}_{\parallel}^{-}(k)\right]_{ij}\sin\phi_{ij},
\end{equation}
\begin{equation}
\label{eq:I2_jQ}
\left[\mathsf{Q}_{\parallel}^{-}(k)\right]_{ij}= \int \vec{Q}^{-}_{ij}(\vec{k})\cdot\hat{k}d\Theta_k\hspace{1.5 cm} \hat{k}=\frac{\vec{k}}{k},
\end{equation}
\begin{equation}
\label{eq:I2_jP}
\left[\mathsf{P}_{\parallel}^{-}(k)\right]_{ij}= \int \vec{P}^{-}_{ij}(\vec{k})\cdot\hat{k}d\Theta_k, \hspace{1.7 cm} \hat{k}=\frac{\vec{k}}{k}.
\end{equation}

The \emph{net radial} component $\mathrm{B}_{\parallel}(k)$ is proportional to the flux of the field $\vec{B}(\vec{k})$, which is responsible for the PECD\cite{ordonez_propensity_2019} [see Eq.
(\ref{eq:conductB})]. Thus, the global invariant $\left[\mathsf{Q}_{\parallel}^{-}(k)\right]_{ij}$ for $i=j$ leads to PECD for photoionization from a real stationary state. PECD from the superposition of states (time-dependent PECD\cite{Comby2016JPCL}) also involves the complementary quadrature $\left[\mathsf{P}_{\parallel}^{-}(k)\right]_{ij}$. 

The \textbf{Class III} of enantio-sensitive observables originates from an infinite array of multipoles ($l\geq 1$) of \emph{net radial} and of the two \emph{net tangential} components of the geometric field: Eqs. (\ref{eq:I3multi_r},\ref{eq:I3multi_t1},\ref{eq:I3multi_t2}). The parity of spherical harmonics $Y_{lm}(\vec{k})=(-1)^lY_{lm}(-\vec{k})$ dictates that even multipoles $\left[\mathsf{B}_{\parallel}^{l=2n,m}(k)\right]_{ij}$ and $\left[\mathsf{B}_{\perp,1}^{l=2n,m}(k)\right]_{ij}$ can only emerge due to the asymmetric quadratures $\vec{Q}^{-}_{ij}(\vec{k})$ and $\vec{P}^{-}_{ij}(\vec{k})$, while the odd multipoles $\left[\mathsf{B}_{\parallel}^{l=2n+1,m}(k)\right]_{ij}$ and $\left[\mathsf{B}_{\perp,1}^{l=2n+1,m}(k)\right]_{ij}$ can emerge only due to the symmetric quadratures $\vec{Q}^{+}_{ij}(\vec{k})$ and $\vec{P}^{+}_{ij}(\vec{k})$. For the $\left[\mathsf{B}_{\perp,2}^{l,m}(k)\right]_{ij}$ multipoles it is the other way around: the terms with even $l$ may only appear due to the symmetric quadratures $\vec{Q}^{+}_{ij}(\vec{k})$ and $\vec{P}^{+}_{ij}(\vec{k})$, while the terms with odd $l$ may emerge only due to $\vec{Q}^{-}_{ij}(\vec{k})$ and $\vec{P}^{-}_{ij}(\vec{k})$. 

Class III observables emerge in two- or multi-photon ionization of randomly oriented molecules. 

For example, quadrupolar [$l=1$ in Eq. ({\ref{eq:I3multi_r}})] PECD currents \cite{ordonez_disentangling_2020} emerge in photoionization of chiral molecules triggered by orthogonally polarized two-color $\omega-2\omega$ fields \cite{Demekhin2018PRL,demekhin_photoelectron_2019,Rozen2019PRX}. Other members of this class have not been identified so far. 

\section{New anomalous enantio-sensitive observables}
 
While the Class II and some of the Class III observables have already been detected in experiments, the Class I of anomalous enantio-sensitive observables have not been explored yet. 
Class I involves any vectorial observable $\vec{V}$ of the cation (see Methods): 
\begin{align}
 \langle\vec{V}^{\mathsf{L}}(k)\rangle&\equiv \int d \rho W (k,\rho)\vec{V}^{\mathsf{L}}(\rho)&\nonumber\\
 &=\frac{1}{2}|{\mathcal{E}}(\omega)|^2\sigma\int d \rho \left[\int \vec{B}^{\mathsf{L}}(\vec{k},\rho)\cdot \hat{z} d \Theta_k \right]\vec{V}^{\mathsf{L}}\nonumber\\
 &=\frac{1}{6}|{\mathcal{E}}(\omega)|^2\sigma\left(\vec{\mathsf{B}}^{\mathsf{M}}(k)\cdot \vec{V}^{\mathsf{M}}\right)\hat{z}.
 \label{global_observable_text}
\end{align}
Here the superscripts $\mathsf{L}$ and $\mathsf{M}$ indicate laboratory or molecular frame vectors, $\rho$ denotes the Euler angles characterizing the orientation of the molecular frame relative to the laboratory frame, $W (k,\rho)$ is the ionization rate into energy $E_k=k^2/2$ for a fixed in space molecule ionised by circularly polarized light.
Equation (\ref{global_observable_text}) shows that after ionization the ensemble-averaged value of any molecular frame vector $\vec{V}^{\mathsf{M}}$ will have an \emph{anomalous} (proportional to the net geometric field) enantio-sensitive component along the direction perpendicular to the polarization plane of the ionizing pulse. $\vec{V}^{\mathsf{M}}$ can represent electronic or vibronic currents, transition dipoles, induced polarization, permanent dipoles. Below we focus on the specific example of an observable quantifying molecular orientation.

Class I observables require excitation of current prior to photoionization.
Consider a pump-probe set-up with two pulses copropagating along the $\hat{z}$ axis of the laboratory frame. The linearly polarized pump pulse excites a coherent superposition of two states with energy difference $\omega_{12}$ in a randomly oriented molecular ensemble. The excitation is probed via photoionization by a circularly polarized probe. The net geometric field is: $\vec{\mathsf{B}}_{12}(k,t)=\vec{\mathsf{P}}^{+}_{12}(k)\sin(\omega_{12}t)$.

Suppose that $\hat{\mathsf{e}}^{\mathsf{M}}_\mathsf{B}$ is a unit polar molecular frame vector (superscript $\mathsf{M}$) collinear with the net geometric field $\hat{\mathsf{e}}^{\mathsf{M}}_\mathsf{B}\parallel\vec{\mathsf{P}}_{12}(k)$ in the molecular frame of a given enantiomer. The scalar product $\hat{\mathsf{e}}^{\mathsf{M}}_\mathsf{B}\cdot\vec{\mathsf{P}}_{12}(k)=\upsilon|\vec{\mathsf{P}}_{12}(k)|$ is a pseudoscalar , which has opposite signs ($\upsilon=\pm1$) in opposite enantiomers. 
Upon photoionization the same vector $\hat{\mathsf{e}}^{\mathsf{L}}_{\mathsf{B}}$ in the laboratory frame (superscript $\mathsf{L}$) is:
\begin{equation}
\label{eq:orientation_full_new}
\langle\hat{\mathsf{e}}_{\mathsf{B}}^{\mathsf{L}}(k,\tau)\rangle = R \langle\hat{\mathsf{e}}_{\mathsf{B}}^{\mathsf{L}}(k,\tau)\rangle_{\mathrm{isotropic}},
\end{equation}
\begin{equation}
\label{eq:orientation_n0_alignment_new}
\langle\hat{\mathsf{e}}_{\mathsf{B}}^{\mathsf{L}}(k,\tau)\rangle_{\mathrm{isotropic}} = \frac{1}{9}C \sigma\upsilon(\vec{d}_{10}\cdot\vec{d}_{20})\left|\vec{\mathsf{P}}^{+}_{12}(k)\right|\sin(\omega_{12}\tau) \hat{z},
\end{equation}
where $R$ accounts for the molecular alignment induced by the pump, 
\begin{equation}
R\equiv\frac{6}{5}\left[1 - \frac{1}{2}\frac{(\vec{d}_{10}\cdot\hat{\mathsf{e}}_\mathsf{B}^{\mathsf{M}})(\vec{d}_{20}\cdot\hat{\mathsf{e}}_\mathsf{B}^{\mathsf{M}})}{(\vec{d}_{10}\cdot\vec{d}_{20})}\right],
\label{eq:orientation_full_factor_new} 
\end{equation} 
while $\langle\hat{\mathsf{e}}_{\mathsf{B}}^{\mathsf{L}}(k,\tau)\rangle_{\mathrm{isotropic}}$ ignores it; $\vec{d}_{10}$ and $\vec{d}_{20}$ are the molecular frame excitation dipoles and $C$ encodes the pump and probe Fourier components at the excitation and photoionization frequencies correspondingly\footnote{We consider transform limited pulses.} (see Methods):
\begin{equation}
C\equiv|\mathcal{E}^{*}_1(\omega_{20})\mathcal{E}_2^*(\omega_{k2})\mathcal{E}_1(\omega_{10})\mathcal{E}_2(\omega_{k1})|.
\label{eq:C_per_dip}
\end{equation}
Eqs. (\ref{eq:orientation_full_new}, \ref{eq:orientation_n0_alignment_new}) show that the oscillating net geometric field $\left|\vec{\mathsf{P}}^{+}_{12}(k)\right|\sin(\omega_{12}\tau)$ dictates the orientation of the molecular frame vector $\hat{{\mathsf{e}}}^{\mathsf{M}}_{{\mathsf{B}}}(k)\parallel\vec{\mathsf{P}}^{+}_{12}(k)$ along the $\hat{z}$-axis perpendicular to the polarization of the circularly polarized probe. The direction of the vector $\hat{\mathsf{e}}_{\mathsf{B}}^{\mathsf{L}}(k,\tau)$ is opposite in opposite enantiomers, because $\vec{\mathsf{P}}^{+}_{12}(k)\cdot \hat{\mathsf{e}}^{\mathsf{M}}_{\mathsf{B}}=\upsilon |\vec{\mathsf{P}}^{+}_{12}(k)|$ is a molecular pseudoscalar that has opposite sign ($\upsilon=\pm1$) in opposite enantiomers. The orientation is also reversed when the direction of rotation ($\sigma=\pm1$) of the circularly polarized probe pulse is reversed. 

Eqs. (\ref{eq:orientation_full_new}, \ref{eq:orientation_n0_alignment_new}) predict that the enantio-sensitive orientation (the molecular orientation circular dichroism, MOCD) oscillates as a function of the pump-probe delay, reaching maximal positive or negative values for $\tau=(2n-1)\pi/(2\omega_{12})$, $n$ is a positive integer. Photoionization at opportune times using a circularly polarized probe induces enantio-sensitive orientation of both the molecular cations and of the excited neutrals that 
where not ionised, with the orientation of the neutrals opposite to that of the cations.

\begin{figure*}[h]
\includegraphics[scale=0.3]{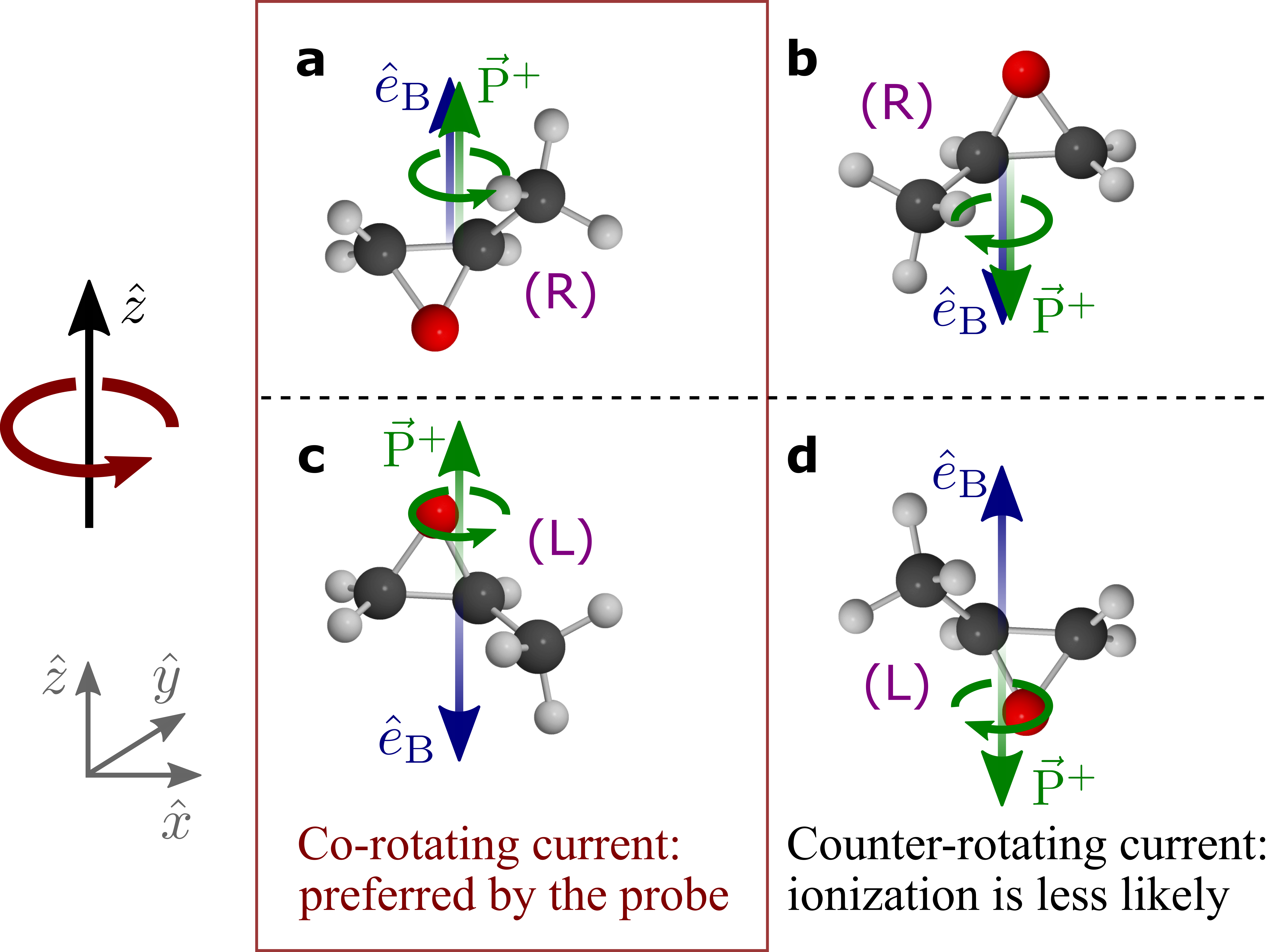}
\caption{\label{fig:Bfield_in_mol} 
Relative orientations of the molecular vectors $\vec{e}_{\mathrm{B}}$ and $\vec{\mathrm{P}}^{+}$ for opposite molecular orientations and opposite enantiomers for $k=0.2$ a.u. and a superposition of LUMO and LUMO+1 in propylene oxide. The red circular arrow shows the rotation direction of the circularly polarized field. The green circular arrows show the circular current in the excited states right before ionization takes place. Photoionization rates are higher for orientations (a) and (c) than for (b) and (d) because ionization is more effective when the electronic current (circular green arrow fixed to $\vec{\mathrm{P}}^{+}$) and the electric field rotate in the same direction. This difference in photoionization rates causes enantio-sensitive orientation. }
\end{figure*}

Figure \ref{fig:Bfield_in_mol} shows the direction of the molecular orientation $\hat{\mathsf{e}}^{\mathsf{L}}_{\mathsf{B}}$ arising when the excitation of LUMO and LUMO+1 in propylene oxide is
followed by photoionization into the states with momentum $k=0.2$ a.u., for the left- and right-handed enantiomers. 
The quadrature $\vec{\mathsf{P}}^{+}_{12}(k)$, where $|2\rangle=|LUMO+1\rangle$, $|1\rangle=|LUMO\rangle$, has the same direction in the left- and right-handed enantiomers, but the pseudoscalar $\upsilon$ has opposite signs for the opposite enantiomers, corresponding to opposite orientations of left and right molecular ions with respect to the laboratory $\hat{z}$ axis. To calculate the geometric field, we have used the photo-excitation and photoionization dipoles computed using the DFT-based approach 
developed in Ref. \cite{Toffoli2002}. This approach yields excellent agreement with the experimental data for one-photon ionization of chiral molecules 
\cite{Turchini2004,Stener2004,Stranges2005,DiTommaso2006,Turchini2009}. 

\begin{figure*}[h]
\includegraphics[width=\textwidth]{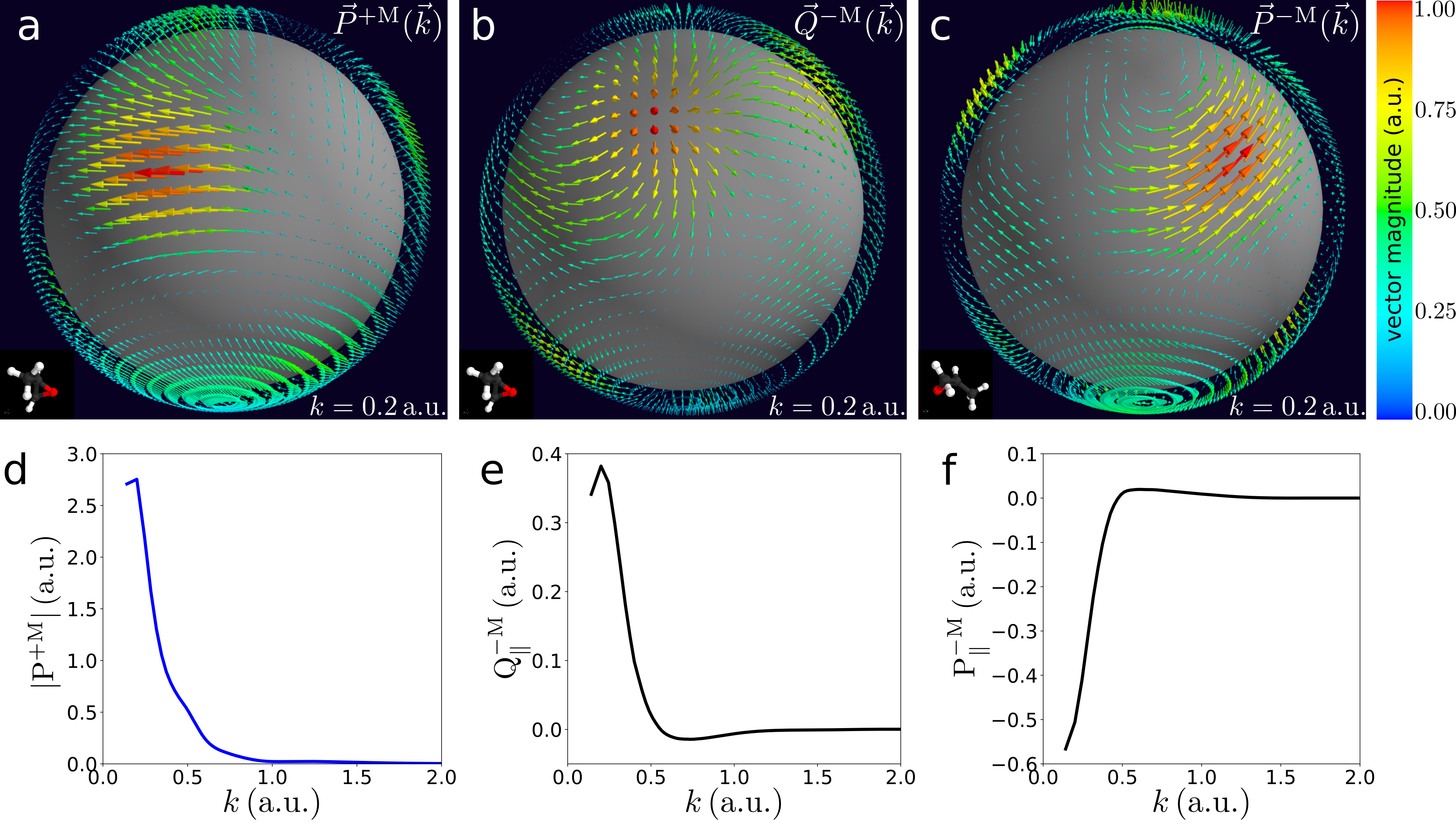} 
\caption{\label{fig:Bfield_comp} Geometric field and its global invariants emerging upon excitation of LUMO and LUMO+1 orbitals in propylene oxide. (a) The symmetric quadrature $\vec{{P}}^{+\mathsf{M}}(\vec{k})$ [Eq. (\ref{eq:P_evenodd_definition})] for $k=0.2$ a.u. and (d) the magnitude of its net value $|\vec{\mathsf{P}}^{+\mathsf{M}}(k)|$ [Eq. (\ref{eq:netB1})], which governs Class I observables, such as enantio-sensitive molecular orientation (MOCD) [Eqs. (\ref{eq:orientation_full_new}, \ref{eq:orientation_n0_alignment_new})].
(b) Asymmetric quadratures $\vec{{Q}}^{-\mathsf{M}}(\vec{k})$ [Eq. (\ref{eq:Q_evenodd_definition})] and (c) $\vec{{P}}^{-\mathsf{M}}(\vec{k})$ [Eq. \ref{eq:P_evenodd_definition}] for $k=0.2$ a.u. and (e) the net values of their radial components $\mathsf{Q}_{\parallel}^{-\mathsf{M}}(k)$ [Eq. (\ref{eq:I2_jQ})] and (f) $\mathsf{P}_{\parallel}^{-\mathsf{M}}(k)$ [Eq. (\ref{eq:I2_jP})], which govern Class II observables, such as the TD-PECD [Eq. (\ref{eq:TDPECD})]}.
\end{figure*}

Figure \ref{fig:Bfield_comp} compares the quadratures responsible for Class I and Class II enantio-sensitive observables for the same excitation. The magnitude of the net quadrature $|\vec{\mathsf{P}}^{+}_{12}(k)|$ quantifying the MOCD is substantially larger than each of the quadratures $\left[\mathsf{Q}_{\parallel}^{-\mathsf{M}}(k)\right]_{12}$ and $\left[\mathsf{P}_{\parallel}^{-\mathsf{M}}(k)\right]_{12}$ quantifying the time-dependent PECD (TD-PECD):
\begin{eqnarray}
\label{eq:TDPECD} 
\vec{j}_{\text{TD-PECD}}^{\mathsf{L}}(k,\tau)= \frac{1}{9}C\sigma k(\vec{d}^{\mathsf{M}}_{10}\cdot\vec{d}^{\mathsf{M}}_{20}) \left\{\cos(\omega_{12}\tau)\left[\mathsf{Q}^{-\mathsf{M}}_{\parallel}(k)\right]_{12}+\sin(\omega_{12}\tau)\left[\mathsf{P}^{-\mathsf{M}}_{\parallel}(k)\right]_{12}\right\}\hat{z}.
\end{eqnarray} 
In Eq. (\ref{eq:TDPECD}) we have ignored the molecular alignment by the pump and omitted the time-independent contributions to facilitate the comparison with Eq. (\ref{eq:orientation_n0_alignment_new}) for MOCD.

Figure \ref{fig:Bfield_comp} shows that the enantio-sensitive orientation is at least of the same order of magnitude as the enantio-sensitive signal in TD-PECD\cite{Comby2016JPCL}. Importantly, TD-PECD and MOCD involve completely different components of the geometric field and therefore expose different and complementary aspects of chiral dynamics in molecules. 

Qualitatively, the orientation induced by photoionization can be understood as follows. First, a linearly polarized pump excites a couple of excited states $|1\rangle$ and $|2\rangle$ and produces a current $\vec{j}_{12}$ oscillating at frequency $\omega_{12}$. 
Suppose that this current goes from the `head' of a molecule to its `tail' in the molecular frame. For two molecules oppositely oriented in the laboratory frame (see Figs. \ref{fig:Bfield_in_mol}a and \ref{fig:Bfield_in_mol}b), $\vec{j}_{12}$ will have the same direction in the molecular frame \footnote{This can be shown by noting that $\vec{j}_{12}$ depends on the product of the transition amplitudes to states $|1\rangle$ and $|2\rangle$}. 
Second, like a helix, the chiral structure of the molecule converts this linear current into a circular current, which will have the same direction of rotation in the molecular frame for both orientations. Thus, in the laboratory frame, the two oppositely oriented molecules will display circular currents rotating in opposite directions (see circular arrows in Figs. \ref{fig:Bfield_in_mol}a and \ref{fig:Bfield_in_mol}b). 
Due to the propensity rules in one-photon ionization, explicitly quantified by the geometric field $\vec{B}_{12}(\vec{k})$, the circularly polarized probe pulse `selects' (or preferentially ionizes) the orientation with the current co-rotating with the probe pulse. This leads to a difference in the photoionization yields resulting from each orientation and thus to the emergence of oriented molecular ions. Note that for the opposite enantiomer, the chiral structure of the molecules creates the opposite circulating current in the molecular frame, and thus the probe-pulse `selects' the opposite orientation (see Figs. \ref{fig:Bfield_in_mol}c and \ref{fig:Bfield_in_mol}d). 

Importantly, the difference in the angle-integrated ionization yields for the two opposite orientations is proportional to the projection of the net geometric field $\vec{\mathrm{B}}_{12}(k)$ on the laboratory $\hat{z}$ axis. Thus, the propensity rule `selecting' molecular orientations with co-rotating current is most pronounced for orientations where $\vec{\mathrm{B}}_{12}(k)\parallel\hat{z}$, which explains why $\vec{\mathrm{B}}_{12}(k)$ is the molecular axis becoming maximally oriented. The fact that the photoionization yield depends on the direction of the 
current excited in the chiral molecule prior to photoionization not only explains the emergence MOCD but is also an example of enantio-sensitive charge directed reactivity.

Figure \ref{fig:orientation} shows the degree of orientation of the molecular ions due to subsequent ionisation by the probe pulse calculated using the DFT-based matrix elements, dipole couplings and Eq. (\ref{eq:orientation_full_new}). The strength of the effect is characterized by the fraction of oriented ions normalized to the averaged over pump-probe delay total ionization rate, $\langle\langle \hat{\mathsf{e}}_{\mathsf{B}}^{\mathsf{L}}(k,\tau)\rangle\rangle$, see Methods. Having $\langle\cos\Theta\rangle=-0.12$ for $k=0.2$ a.u. means that roughly 59 out of 100 molecules have $\hat{\mathsf{e}}_\mathsf{B}\cdot\hat{z}<0$ and 41 out of 100 molecules have $\hat{\mathsf{e}}_\mathsf{B}\cdot\hat{z}>0$. Thus, a very significant degree 
of enantio-sensitive orientation results from 
the excitation of chiral dynamics in valence shells.

\section{Outlook}

We see several future directions associated with geometric fields in photoionization of chiral molecules.

Valence-shell MOCD can be induced by exciting electronic or vibronic degrees of freedom and thus can be achieved with pulses of various durations: from sub-femtoseconds to picoseconds. 
An interesting future direction is to use MOCD, which occurs both in molecular ions and excited neutral molecules remaining after ionization, for (i) quantification of helical currents in chiral molecules, (ii) enantio-separation and (iii) ultrafast molecular imaging with oriented chiral molecules. 

MOCD can also be induced by core excitation with few-femtosecond X-ray pulses. Localised site-specific core excitation could allow one to initiate currents from different locations inside the molecule and probe the orientation of the most efficient ``molecular cork-screw''.

Probing the induced electronic excitations 
should likely be done before the core-hole decay due to e.g. the Auger process. Molecular fragmentation, which may be induced by such decay, could be beneficial for detecting MOCD via
angular distributions of fragments.

We have found that MOCD can also be induced using circularly polarized pump pulse and linearly polarized probe. In this case, MOCD emerges due to the helical PXCD current \cite{beaulieu_photoexcitation_2018} excited in bound states \cite{Ordonez_Smirnova_prep} and does not involve geometry of photoionization dipoles. MOCD induced by circularly polarized pump and probe pulses records the interference between the PXCD mechanism of MOCD\cite{Ordonez_Smirnova_prep} and the geometric mechanism described here. Such interference presents a new opportunity to characterize the geometric field experimentally.

 Direct experimental detection of the geometric field $\vec{B}(\vec{k})$ requires measurement of circular dichroism in the photoionization yield for a given photoelectron momentum, resolved on the orientation of molecular frame, for all possible orientations. Experiments\cite{PhysRevLett.126.083201} indicate that such measurement is within the reach for small size chiral molecules.

Identification of new members of Classes I, II, and III of enantio-sensitive observables predicted in this work is another exciting new direction. 

The geometric field
in chiral molecules, described here, is a new member of the family of geometric fields. It should allow one to explore the interplay between chiral and topological phenomena in chiral molecules and solids.

\section{Acknowledgments}
We gratefully acknowledge many enlightening discussions with Prof. Misha Ivanov. We thank Dr. Rui Emanuel Ferreira da Silva, Dr. Álvaro Jiménez Galán, Dr. Kiyoshi Ueda for their comments on the manuscript and Dr. Emilio Pisanty for suggesting the acronym "MOCD". OS is grateful to Ms. Julia Riedel for applying right pressure at right times.
O.S., D.A., A.F.O. acknowledge funding from Deutsche Forschungsgemeinschaft (SM 292/5-2).
A.F.O. acknowledges grants supporting research at ICFO: Agencia Estatal de Investigación (“Severo Ochoa” Center of Excellence CEX2019-000910-S, National Plan FIDEUA PID2019-106901GB-I00/10.13039 / 501100011033, FPI), Fundació Privada Cellex, Fundació Mir-Puig, Generalitat de Catalunya (AGAUR Grant No. 2017 SGR 1341, CERCA program) and EU Horizon 2020 Marie Skłodowska-Curie grant agreement No 101029393, supporting his research on chirality. D.A. acknowledges funding from Royal Society (URF/R1/201333) supporting his research on chirality.

\section{Methods}
\subsection{Operator approach and time-reversal symmetry of the net geometric field}
Here we prove that
\begin{equation}
 \int\mathrm{d}\Theta_{k}\vec{Q}_{ij}(\vec{k})=0
\end{equation}
when $|\psi_i\rangle$ and $|\psi_j\rangle$ are time-even states, which we use in Eq. (\ref{eq:netB1}). 

The time-reversal symmetry plays a central role in enabling the net geometric field. To consider the parity of the geometric field (or its quadratures) with respect to time-reversal, it is convenient\cite{suzuki2018} to introduce an operator approach. Eq. (\ref{eq:B_ij}) of the main text can be rewritten in the following form:
\begin{equation}
\label{eq:B_ij_app}
\vec{B}_{ij}(\vec{k},\phi_{ij})=-\frac{1}{2}i\left[\vec{d}^{*}_{\vec{k}i}\times \vec{d}_{\vec{k}j}\right]e^{i\phi_{ij}}+\mathrm{c.c.}=-\frac{1}{2}\vec{F}_{ij}e^{i\phi_{ij}}+\mathrm{c.c.},
\end{equation}
where $\vec{F}_{ij}(\vec{k})\equiv i\left[\vec{d}^{*}_{\vec{k}i}\times \vec{d}_{\vec{k}j}\right]$ and thus,
\begin{equation}
\vec{Q}_{ij}(\vec{k})\equiv -\Re
\left\{\vec{F}_{ij}(\vec{k})\right\},
\label{eq:Q_ij_definition_app}
\end{equation}
\begin{equation}
\vec{P}_{ij}(\vec{k})\equiv \Im\left\{ \vec{F}_{ij}(\vec{k})\right\}.
\label{eq:P_ij_definition_app}
\end{equation}
To establish the parity of net quadratures with respect to time-reversal, it is sufficient to consider the time-reversal parity of the net $\vec{F}_{ij}(\vec{k})$:
\begin{align}
\vec{\mathrm{F}}_{ij} & \equiv\int\mathrm{d}\Theta_{k}i\vec{d}_{\vec{k},i}^{*}\times\vec{d}_{\vec{k},j}.
\end{align}
Analogously to the procedure in Ref. \cite{suzuki2018}, we rewrite $\vec{\mathrm{F}}_{ij}$ as the matrix element
of an operator $\hat{\vec{\mathrm{F}}}$:
\begin{align}
\vec{\mathrm{F}}_{ij} & \equiv\int\mathrm{d}\Theta_{k}i\vec{d}_{\vec{k},i}^{*}\times\vec{d}_{\vec{k},j}=\int\mathrm{d}\Theta_{k}i\langle\psi_{i}|\hat{\vec{r}}|\psi_{\vec{k}}^{\left(-\right)}\rangle\times\langle\psi_{\vec{k}}^{\left(-\right)}|\hat{\vec{r}}|\psi_{j}\rangle=\langle\psi_{i}|\hat{\vec{\mathrm{F}}}|\psi_{j}\rangle,
\end{align}
where the operator $\hat{\vec{\mathrm{F}}}$ in vector and component
form is given by
\begin{align}
\hat{\vec{\mathrm{F}}} & \equiv i\int\mathrm{d}\Theta_{k}\hat{\vec{r}}|\psi_{\vec{k}}^{\left(-\right)}\rangle\times\langle\psi_{\vec{k}}^{\left(-\right)}|\hat{\vec{r}};\qquad\hat{\mathrm{F}}_{a}=i\epsilon_{abc}\int\mathrm{d}\Theta_{k}\hat{r}_{b}\hat{P}_{\vec{k}}\hat{r}_{c}=i\epsilon_{abc}\hat{r}_{b}\hat{P}_{k}\hat{r}_{c},
\end{align}
and $\hat{P}_{k}\equiv\int\mathrm{d}\Theta_{k}|\psi_{\vec{k}}^{\left(-\right)}\rangle\langle\psi_{\vec{k}}^{\left(-\right)}|$
is the projector on the subspace of all states with energy $E=k^{2}/2$.
Since $\hat{\vec{r}}$ and $\hat{P}_{k}$ are Hermitian, $\hat{\vec{\mathrm{F}}}$ is also Hermitian:
\begin{align}
\hat{\mathrm{F}}_{a}^{\dagger} & =\left(i\epsilon_{abc}\hat{r}_{b}\hat{P}_{k}\hat{r}_{c}\right)^{\dagger}=-i\epsilon_{abc}\hat{r}_{c}\hat{P}_{k}\hat{r}_{b}=i\epsilon_{acb}\hat{r}_{c}\hat{P}_{\vec{k}}\hat{r}_{b}=\hat{\mathrm{F}}_{a}.
\end{align}

Now we want to see how time reversal considerations constraint the
matrix elements of $\hat{\vec{\mathrm{F}}}$. Let $\hat{T}$ be the
time-reversal operator (see e.g. sec. 4.4 in Ref. {\cite{sakurai1995}}). $\hat{T}$ converts a state $|\alpha\rangle$
into its time-reversed counterpart $|\tilde{\alpha}\rangle=\hat{T}|\alpha\rangle$,
which we denote by a tilde. $\hat{T}$ is anti-unitary, i.e. $\langle\tilde{\alpha}|\tilde{\beta}\rangle=\langle\alpha|\beta\rangle^{*}$
and $T\left(c_{1}|\alpha\rangle+c_{2}|\beta\rangle\right)=c_{1}^{*}|\tilde{\alpha}\rangle+c_{2}^{*}|\tilde{\beta}\rangle$.
A Hermitian operator $\hat{A}$ is time-even ($+$) or time-odd ($-$)
when $\hat{T}\hat{A}=\pm\hat{A}\hat{T}$. Time parity restricts the matrix elements
of $\hat{A}$ according to $\langle\alpha|\hat{A}|\beta\rangle=\pm\langle\tilde{\alpha}|\hat{A}|\tilde{\beta}\rangle^{*}.$
In particular, if $\hat{A}$ is time-odd and the states are time-even
(i.e. $|\tilde{\alpha}\rangle=|\alpha\rangle$ and $|\tilde{\beta}\rangle=|\beta\rangle$),
it follows that $\langle\alpha|\hat{A}|\beta\rangle=-\langle\alpha|\hat{A}|\beta\rangle^{*}$ and thus $\Re\{\langle\alpha|\hat{A}|\beta\rangle\}=0$.

For $\hat{P}_{k}$ we have
\begin{align}
\hat{T}\hat{P}_{k}|\phi\rangle & =\hat{T}\left[\int\mathrm{d}\Theta_{k}|\psi_{\vec{k}}^{\left(-\right)}\rangle\langle\psi_{\vec{k}}^{\left(-\right)}|\phi\rangle\right]=\int\mathrm{d}\Theta_{k}|\tilde{\psi}_{\vec{k}}^{\left(-\right)}\rangle\langle\psi_{\vec{k}}^{\left(-\right)}|\phi\rangle^{*}=\int\mathrm{d}\Theta_{k}|\tilde{\psi}_{\vec{k}}^{\left(-\right)}\rangle\langle\tilde{\psi}_{\vec{k}}^{\left(-\right)}|\tilde{\phi}\rangle=\hat{\tilde{P}}_{k}\hat{T}|\phi\rangle
\end{align}
where $\hat{\tilde{P}}_{k}\equiv\int\mathrm{d}\Theta_{k}|\tilde{\psi}_{\vec{k}}^{\left(-\right)}\rangle\langle\tilde{\psi}_{\vec{k}}^{\left(-\right)}|$
also projects on the subspace of all states with energy $E=k^{2}/2$. Since $\hat{P}_{k}$ and $\hat{\tilde{P}}_{k}$ project on the same
subspace we must have $\hat{P}_{k}=\hat{\tilde{P}}_{k}$ and thus $\hat{P}_{k}$ is time-even. Furthermore, since $\hat{\vec{r}}$ is also time-even, we can use the anti-unitary character of $\hat{T}$
to find that
\begin{equation}
\hat{T}\hat{\mathrm{F}}_{a}|\phi\rangle=\hat{T}\left(i\epsilon_{abc}\hat{r}_{b}\hat{P}_{k}\hat{r}_{c}|\phi\rangle\right)=-i\epsilon_{abc}\hat{r}_{b}\hat{P}_{k}\hat{r}_{c}\hat{T}|\phi\rangle=-\hat{\mathrm{F}}_{a}\hat{T}|\phi\rangle,
\end{equation}
which means that $\hat{\vec{\mathrm{F}}}$ is time-odd. Thus, $\Re\{\vec{\mathrm{F}}_{ij}\}=0$
when $|\psi_{i}\rangle$ and $|\psi_{j}\rangle$
are time-even states. Finally, using Eqs. (\ref{eq:Q_ij_definition_app},\ref{eq:P_ij_definition_app})
we obtain for time-even states $|\psi_{i}\rangle$ and $|\psi_{j}\rangle$:
\begin{equation}
\int\mathrm{d}\Theta_{k}\vec{Q}_{ij}(\vec{k})=-\int\mathrm{d}\Theta_{k}\Re\left\{ \vec{F}_{ij}(\vec{k})\right\} =-\Re\left\{ \vec{\mathrm{F}}_{ij}\right\} =0,
\end{equation}

Further, the time-odd parity of $\hat{\vec{\mathrm{F}}}$ does not restricts the value of net momentum quadrature, which in general is non-zero:
\begin{equation}
\int\mathrm{d}\Theta_{k}\vec{P}_{ij}(\vec{k})=\int\mathrm{d}\Theta_{k}\Im\left\{\vec{F}_{ij}(\vec{k})\right\} =\Im\left\{ \vec{\mathrm{F}}_{ij}\right\} \neq0.
\end{equation}

\subsection{The origin of Class I enantio-sensitive observables}
The connection between enantio-sensitive observables in photoionization and the geometric field becomes evident as soon as we consider the photoionization rate $W(k)$ into energy $E=k^2/2$ for a fixed in space molecule ionised by circularly polarized light 
in the electric dipole approximation \cite{ordonez_generalized_2018}:
\begin{equation}
 W(k,\rho)\propto\int d \Theta_k |a_{\vec{k}c}|^2
\end{equation}
\begin{equation}
 |a_{\vec{k}c}|^2=\frac{1}{2}|{\mathcal{E}}(\omega)|^2\left\{|\vec{d}^{\mathsf{L}}_{\vec{k}c}(\rho)\cdot\hat{x}|^2+|\vec{d}^{\mathsf{L}}_{\vec{k}c}(\rho)\cdot\hat{y}|^2+\sigma\vec{B}^{\mathsf{L}}(\vec{k},\rho)\cdot\hat{z}\right\}.
 \label{yield}
\end{equation}
Here $a_{\vec{k}c}$ is the amplitude of ionization from a complex-valued randomly oriented bound state $|\psi_{c}\rangle$ (at this point we do not specify how this state was created in a randomly oriented molecular ensemble and assume that it is given to us) into the final state with photoelectron momentum $\vec{k}$, $\mathcal{E}(\omega)$ is the Fourier component of the light field at the transition frequency $\omega$ \footnote{We define the Fourier transform as $\tilde{\vec{E}}(\omega)=\int_{-\infty}^{\infty}\vec{E}(t)e^{i\omega t}dt$ and assume $\tilde{\vec{E}}(\omega)=\mathcal{E}(\omega)(\vec{x}-i\sigma\vec{y})/\sqrt{2}$ at the transition frequency}, $\sigma=\pm1$ for light rotating clockwise/counterclockwise in the $xy$ plane, integration over $d \Theta_k$ describes averaging over the directions of the photoelectron momentum, $\rho$ denotes the Euler angles characterizing the orientation of the molecular frame relative to the laboratory frame, the vectors $\hat{x}$, $\hat{y}$, and $\hat{z}$ denote the axes of the laboratory frame, $\vec{d}^{\mathsf{L}}_{\vec{k}c}(\rho)$ is the photoionization dipole in the laboratory frame (denoted by superscript ${\mathsf{L}}$), and $\vec{B}^{\mathsf{L}}(\vec{k},\rho)$ is the geometric field in the laboratory frame.

Using Eq. (\ref{yield}) we formally obtain the expression for the orientation-averaged value of an arbitrary vectorial observable $\vec{V}^{\mathsf{L}}(k)$ of the molecular cation:
\begin{align}
 \langle\vec{V}^{\mathsf{L}}(k)\rangle&\equiv \int d \rho W (k,\rho)\vec{V}^{\mathsf{L}}(\rho)&\nonumber\\
 &=\frac{1}{2}|{\mathcal{E}}(\omega)|^2\sigma\int d \rho \left[\int \vec{B}^{\mathsf{L}}(\vec{k},\rho)\cdot \hat{z} d \Theta_k \right]\vec{V}^{\mathsf{L}}\nonumber\\
 &=\frac{1}{6}|{\mathcal{E}}(\omega)|^2\sigma\left(\vec{\mathsf{B}}^{\mathsf{M}}(k)\cdot \vec{V}^{\mathsf{M}}\right)\hat{z}.
 \label{global_observable}
\end{align}
Here the superscripts $\mathsf{L}$ and $\mathsf{M}$ indicate that the respective vectors are expressed with respect to the laboratory frame or the molecular frame correspondingly. 
Equation (\ref{global_observable}) shows that after ionization with circularly polarized light, the ensemble-averaged value of the vector $V$ (fixed in the molecular frame), will have an \emph{anomalous} (proportional to the geometric field) enantio-sensitive component along the direction perpendicular to the polarization plane. 

\subsection{Equations describing the MOCD}
Suppose that $\hat{\mathsf{e}}^{\mathsf{M}}_\mathsf{B}$ is a unit polar vector collinear with the net geometric field $\hat{\mathsf{e}}^{\mathsf{M}}_\mathsf{B}\parallel\vec{\mathsf{B}}^{\mathsf{M}}(k)$ in the molecular frame of a given enantiomer. The scalar product $\hat{\mathsf{e}}^{\mathsf{M}}_\mathsf{B}\cdot\vec{\mathsf{B}}^{\mathsf{M}}(k)=\upsilon|\vec{\mathsf{B}}^{\mathsf{M}}(k)|$ is a pseudoscalar ($\upsilon=\pm1$), which has opposite signs in opposite enantiomers. The orientation of the vector $\hat{\mathsf{e}}^{\mathsf{M}}_\mathsf{B}$ in the laboratory frame $\hat{\mathsf{e}}^{\mathsf{L}}_\mathsf{B}$ is given by [see Eq. (\ref{global_observable})]:
 \begin{equation}
 \langle\hat{\mathsf{e}}_\mathsf{B}^{\mathsf{L}}(k)\rangle=\frac{1}{6}|{\mathcal{E}}(\omega)|^2\sigma\upsilon|\vec{\mathsf{B}}^{\mathsf{M}}(k)|\hat{{z}}.
 \label{global_observable1}
\end{equation}
Therefore, Eqs. (\ref{global_observable}) and (\ref{global_observable1}) predict enantio-sensitive orientation of molecular ions by ionization. 
The molecular-frame vector $\hat{\mathsf{e}}^{\mathsf{M}}_\mathsf{B}(k)\parallel\vec{\mathsf{B}}^{\mathsf{M}}(k)$ gets oriented along the laboratory $\hat{{z}}$-axis (perpendicular to the polarization of the circularly polarized probe). 

Now we can specify the procedure of exciting the state $|\psi_{c}\rangle$ in randomly oriented molecular ensemble. The complex-valued state $|\psi_{c}\rangle$ corresponds to excitation of complex superposition of states prior to photoionization, which can be excited with a linearly polarized pump pulse.

Using our approach \cite{ordonez_molecular_2020,ordonez_generalized_2018} we can calculate the orientation of the vector $\hat{\mathsf{e}}^{\mathsf{L}}_{\mathsf{B}}$ in a molecular cation analytically for the excitation of two intermediate states with energy difference $\omega_{2,1}$ 
in an ensemble of randomly oriented chiral molecules. 
Eqs. (\ref{eq:orientation_full_new},\ref{eq:orientation_n0_alignment_new}) of the main text can be obtained using Eqs. (30)-(35) in Ref. \cite{ordonez_generalized_2018} and replacing $\vec{k}$ by $\hat{\mathsf{e}}_{\mathsf{B}}$, choosing a pump linearly polarized along either $\hat{x}$, or $\hat{y}$, and a probe circularly polarized in the $xy$ plane (both pulses are transform limited). To express all resulting terms via the vector product of two ionization dipoles, we used the Binet-Cauchy identity $(\vec{A}\times\vec{B})\cdot(\vec{C}\times\vec{D})=(\vec{A}\cdot\vec{C})(\vec{B}\cdot\vec{D}) - (\vec{A}\cdot\vec{D})(\vec{B}\cdot\vec{C})$ and the vector triple product identity $\vec{A}\times(\vec{B}\times\vec{C})=(\vec{A}\cdot\vec{C})\vec{B} - (\vec{A}\cdot\vec{B})\vec{C}$.

Eq. (26) of the main text can be obtained using Eqs. (30)-(35) in Ref. \cite{ordonez_generalized_2018}, choosing a pump linearly polarized along either $\hat{x}$, or $\hat{y}$, and a probe circularly polarized in the $xy$ plane (both pulses are transform limited).

\begin{figure*}[h]
\center
\includegraphics[scale=0.5]{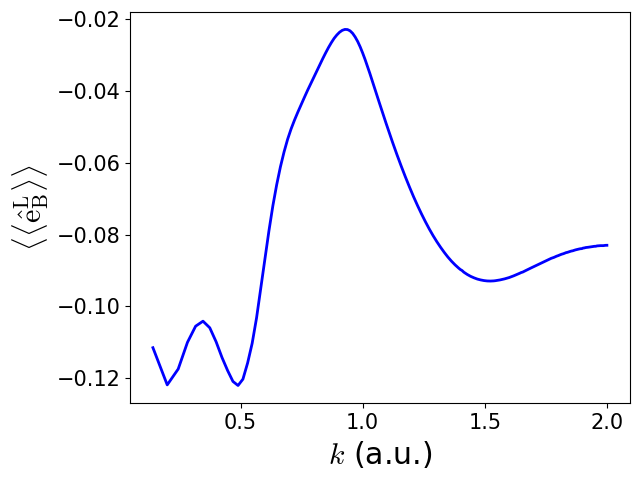}
\caption{\label{fig:orientation}
Degree of orientation $\langle\langle \hat{\mathsf{e}}_{\mathsf{B}}^{\mathsf{L}}(k,\tau)\rangle\rangle$ [see Eq. (\ref{global_observablek}) in Methods] corresponding to coherent excitation of LUMO and LUMO+1 in randomly oriented propylene oxide followed by ionization by circularly polarized pulse, for $\omega_{12}\tau=\pi/2$, $\upsilon$=1, $\sigma=1$.}
\end{figure*}

The strength of the effect is characterized by the fraction of oriented ions normalized to total ionization yield (averaged over pump-probe delay). 
 \begin{equation}
 \langle\langle \hat{\mathsf{e}}_{\mathsf{B}}^{\mathsf{L}}(k,\tau)\rangle\rangle\equiv \frac{\int d \rho \hat{\mathsf{e}}_{\mathsf{B}}^{\mathsf{L}}(\rho) W (k,\rho)}{\int d \rho \overline{W} (k,\rho)},
 \label{global_observablek}
\end{equation} 
\begin{equation}
 \int d \rho \overline{W} (k,\rho) =\sum_{i=1,2} \frac{|C_i|^2}{30}\int\mathrm{d}\Theta_{k}\left[3\left|\vec{d}^{\mathsf{M}}_{i0}\right|^{2}\left|\vec{d}^{\mathsf{M}}_{\vec{k}i}\right|^{2}-\left|\vec{d}^{\mathsf{M}}_{i0}\cdot\vec{d}^{\mathsf{M}}_{\vec{k}i}\right|^{2}\right],
\end{equation}
and $C_i\equiv-\mathcal{E}(\omega_{i0})\mathcal{E}(\omega_{ki})$. These equations are used to calculate the input for Fig. \ref{fig:orientation}.

An estimate of the number of `up' $N_+$ and `down' $N_-$ molecules for $\langle\cos\Theta\rangle=-0.12$ can be performed using a simple model for the angular distribution of oriented molecules: $\Psi(\Theta)=a_0Y_{00}(\Theta)+a_1Y_{10}(\Theta)$, where $a_0^2+a_1^2=1$. Then $\langle\cos\Theta\rangle=\int_{0}^{\pi}d\Theta \int_{0}^{2\pi} d\Phi \sin\Theta\cos\Theta|\Psi(\Theta)|^2$, and we can obtain $N_+\equiv\int_{0}^{\pi/2} d\Theta\int_{0}^{2\pi} d\Phi \sin\Theta|\Psi(\Theta)|^2$ $=\frac{1}{4}(2+3\langle\cos\Theta\rangle)$ $=0.41$ and $N_-\equiv\int_{\pi/2}^{\pi}d\Theta \int_{0}^{2\pi}d\Phi \sin\Theta|\Psi(\Theta)|^2$ $=1-N_+$ $=0.59$.

\bibliography{Bibliography}

\newpage

\end{document}